\documentclass[preprint,aps,tightenlines,showkeys,nofootinbib,superscriptaddress,amsmath]{revtex4}
\usepackage{latexsym}
\usepackage{graphicx}
\usepackage{xcolor}
\newcommand{\beq}{
\begin{eqnarray}}
  \newcommand{\eeq}{
\end{eqnarray}}
\newcommand{\bea}{
\begin{eqnarray*}}
  \newcommand{\eea}{
\end{eqnarray*}}
\newcommand{\eq}{eqnarray}

\newcommand{\ga}{{\gamma}}

\newcommand{\ep}{{\epsilon}}

\newcommand{\De}{\Delta}
\newcommand{\ka}{\kappa}

\newcommand{\la}{{\lambda}}
\newcommand{\La}{{\Lambda}}

\newcommand{\si}{{\sigma}}
\newcommand{\Si}{{\Sigma}}

\newcommand{\Om}{{\Omega}}
\newcommand{\pa}{{\partial}}
\newcommand{\no}{{\nonumber}}
\newcommand{\f}{\frac}

\newcommand{\Ho}{Ho\v{r}ava}

\newcommand{\DiffF}{${\it Diff}_{\cal F}$}

\begin{document}

\preprint{arXiv:2611.30337v2 [hep-th]}

\title{
  Rotating Black Holes from Scaling {Transformations} in \Ho~
  Gravity
}

\author{Deniz O. Devecio\u{g}lu \footnote{E-mail address: dodeve@gmail.com}}
\affiliation{Daegu Gyeongbuk Institute of Science and
  Technology (DGIST), 333 Techno Jungang-daero,
Dalseong-gun, Daegu 42988  Korea }

\author{Mu-In Park \footnote{E-mail address: muinpark@gmail.com,
Corresponding author}}
\affiliation{ Center for Quantum Spacetime, Sogang University,
Seoul, 121-742, Korea }

\date{\today}

\begin{abstract}
  {We study the scaling transformations that change the
    physical parameters of
  solutions in low-energy \Ho~ gravity. In particular, we present}
  a simple scaling
  {transformation that generates} {rotating
  black-hole solutions in} low{-}energy
  \Ho~gravity
  from
  known relativistic solutions.
  {As explicit examples, we obtain {\it ``Doran-like"} and
    {\it ``Kerr-Schild-like"}
    rotating solutions in \Ho~gravity. We find that both solutions are
    {\it horizon-penetrating}, but their null horizons
    are not Killing horizons. We also find that, although their
    singularity structures differ from each other,
    chronology is protected in both solutions,
    as in the previously obtained Kerr-like solutions in
  Boyer-Lindquist coordinates.}
\end{abstract}

\maketitle

\section{Introduction}

With the recent advent of gravitational wave observations
\cite{LIGO:2016} and black-hole imaging by the Event Horizon
Telescope \cite{EHT:2019}, black holes have become key laboratories for
testing Einstein's general relativity (GR) and
modified theories of gravity beyond GR. In particular, obtaining
exact non-Kerr rotating stationary black holes in
modified gravity theories is
essential for such tests, since they allow concrete
comparisons between theoretical predictions and observations.

In this paper, we focus on \Ho~gravity
which has been proposed as a power-counting
renormalizable theory of quantum gravity
free of the ghost problem
by abandoning Lorentz symmetry \cite{DeWi:1967,Hora:2009}. In particular, we
study rotating black-hole solutions in the low-energy
sector of \Ho~gravity and show that exact rotating black-hole
solutions can be generated by scaling transformations that change
their physical parameters.
In Sec. II, we study these scaling transformations and discuss a
simple scaling transformation that generates rotating
black-hole solutions in low-energy \Ho~gravity from relativistic
solutions. In Sec. III, we study
{\it ``Doran-like"} and {\it ``Kerr-Schild-like"}
rotating solutions in
\Ho~gravity as explicit examples. In Sec. IV, we study
their unusual horizon and singularity structures, as well as their
crucial roles in
chronology protection.

\section{Lapse-Shift Scaling Transformations}
In \Ho's formulation of gravity
\cite{Hora:2009}, the time
dimension plays a privileged role. This is achieved by considering a foliation
of the manifold $\mathcal{M}$, where the leaves of foliation
are
constant-time hypersurfaces{\it , i.e.,}
$\mathcal{M}= \bf{R}\times
\Sigma_t$. The foliation is characterized by the
future-directed unit normal vector
\footnote{Throughout this work, Greek indices $\mu,\nu, \cdots
  =0,\cdots,3$ denote spacetime coordinates, while lowercase Latin
  indices $i,j,\cdots=1,\cdots,3$ denote spatial coordinates on the
hypersurfaces $\Sigma_t$.}
\begin{equation}
  n^{\mu }= \left({\frac{1}{N},-\frac{N^{i}}{N}}\right),
\end{equation}
where the lapse $N$ measures how much proper time elapses between
neighboring slices. The shift vector ${N^i}$ determines
how the spatial
coordinates shift from one slice to the next. Together with the
induced metric $g_{ij}$ on $\Sigma_t$, these variables constitute the ADM
decomposition of spacetime
\begin{\eq}
  \label{metric}
  ds^2=-N^2  dt^2+g_{ij}\left(dx^i+N^i dt\right)\left(dx^j+N^j
  dt\right).
\end{\eq}

In what follows, we
consider the low-energy sector of Ho\v{r}ava
gravity \cite{DeWi:1967,Hora:2009,Blas:2009}
described by the action, up to
boundary terms,
\begin{\eq}
  S_g &= & \int_{{\bf R} \times \Si_t} dt d^3 x
  \sqrt{g}N\left[\frac{1}{\kappa}\left(K_{ij}K^{ij}-\lambda
  K^2\right)+\xi R^{}-2 \La {+\frac{\sigma}{2}a_i a^i}\right]\ ,
  \label{horava}
\end{\eq}
where
$
K_{ij}=({2N})^{-1}\left(\dot{g}_{ij}-\nabla_i
N_j-\nabla_jN_i\right)
$
is the extrinsic curvature [the overdot $(\dot{})$ denotes the time
  derivative, 
  $\nabla_i$ is the covariant derivative
  with respect to the
  induced metric $g_{ij}$, and 
  $R$ is the corresponding three-curvature
scalar] and $a_i=\nabla_iN/N$ is the proper acceleration.
The action is invariant under ``foliation-preserving"
diffeomorphisms (\DiffF)
\cite{Hora:2009,Park:2009}. In the infrared limit, for the
general-relativistic
values of the couplings{,
  $\lambda=1,
\kappa \xi=1$ (depending on conventions), and $\sigma=0$,} the
theory reproduces
Einstein gravity.

Taking the ADM variables $(N, N_i, g_{ij})$ as
the fundamental dynamical variables, we
have the following equations of motion (see the
Appendix for details)
\beq
{\cal H}&\equiv&\f{1}{\kappa}\left(K_{ij}K^{ij} -\lambda
K^2\right)-\xi R +2 \La
{-\si \left(\frac{1}{2}\frac{\nabla_i N \nabla^i N}{N^2}-\frac{\nabla_k
\nabla^k N}{N} \right)=0}, \\
{\cal H}^i&\equiv&{-}\f{2}{\ka}\nabla_j
\left(K^{ji}-\lambda\,Kg^{ji}\right)=0\, ,\\
E_{ij}&\equiv&\frac{1}{\kappa}\left( E_{ij}^{(1)}-\lambda E_{ij}^{(2)} \right)
+\xi E_{ij}^{(3)}+\La N g_{ij}+\frac{\si}{2} E_{ij}^{(4)}=0 .
\eeq

Motivated by the preferred role of the foliation in Hořava gravity,
we consider a class of deformations that preserve the intrinsic
geometry of the spatial hypersurfaces while modifying their embedding
into spacetime{, {\it i.e.,} modifying the extrinsic
geometry of the foliation}. Specifically, we change
the ratio of $N_i / N$ while keeping the three-geometry
$g_{ij}$ the same such that
only the extrinsic geometry is modified.

To this end, we
consider the scaling transformations on the lapse and
shift vector
\begin{eqnarray}
  N\rightarrow \alpha  N, \quad N_i &\rightarrow &\beta  N_i.
\end{eqnarray}
These transformations lead
to an extrinsic curvature scaling
\begin{eqnarray}
  K_{ij} &\rightarrow &\frac{\beta}{\alpha } K_{ij},
\end{eqnarray}
provided the metric is stationary, $\partial_t g_{ij}=0$. It is
important to emphasize that these transformations are not
diffeomorphisms unless $\alpha =\beta$.

Applying these transformations to
the field equations,  we find
\begin{eqnarray}
  {\cal H}\rightarrow {\cal H} ^{'}  &=&\f{1
  }{\kappa}\left(\frac{\beta}{\alpha
  }\right)^{2}
  \left[K_{ij}K^{ij} -\lambda
    K^2 -\kappa \xi \left(\frac{\alpha}{\beta}\right)^2
    R+{ 2\kappa \Lambda
    \left(\frac{\alpha}{\beta}\right)^2 }\right. \nonumber \\
    && ~~~~~~~~~~~~~~{\left.- \kappa \si
      \left(\frac{\alpha}{\beta}\right)^2
      \left(\frac{1}{2}\frac{\nabla_i N \nabla^i N}{N^2}-\frac{\nabla_k
  \nabla^k N}{N} \right) \right]=0} ,\label{eom1} \\
  {\cal H}^i\rightarrow {\cal H'}^{i}&=&{-}\f{1}{\ka}\left(\frac{\beta}{\alpha
  }\right)
  \nabla_j
  \left(K^{ji}-\lambda\,Kg^{ji}\right){=0}\, ,\label{eom2}\\
  E_{ij}\rightarrow E'_{ij}&=&\frac{1}{\kappa}
  \left(\frac{\beta^2}{\alpha }\right)
  \left[
    E_{ij}^{(1)}-\lambda
    E_{ij}^{(2)} +\kappa \xi \left(\frac{\alpha}{\beta}\right)^2 E_{ij}^{(3)}
    +\kappa \Lambda \left(\frac{\alpha}{\beta}\right)^2 N
    g_{ij}  \right. \nonumber \\
    && ~~~~~~~~~~~~~~\left.+\frac{\kappa \si}{2}
  \left(\frac{\alpha}{\beta}\right)^2 E_{ij}^{(4)}\right]=0. \label{eom3}
\end{eqnarray}

It is now straightforward to see that one can generate
another
theory with new coupling constants
$\kappa', \xi', \La',\si'$, satisfying
$\kappa' \xi'=\kappa \xi (\alpha/\beta)^2,
\kappa' \Lambda'=\kappa \Lambda (\alpha/\beta)^2,
\kappa' \si'=\kappa \si (\alpha/\beta)^2$,
but with the same
$\la$, from a
solution of low-energy \Ho~gravity (\ref{horava})
(or
vice versa). In particular, if one starts from a solution of
the theory with coupling constants $\kappa', \xi', \La',\si',
\la'$, satisfying
\beq
\kappa' \xi'=\kappa \xi \left(\frac{\alpha}{\beta}\right)^2
\equiv 1,~ \frac{ \Lambda'}{\xi'}=\frac{ \Lambda}{\xi},~ \frac{
\si'}{\xi'}=\frac{ \si}{\xi}, ~\la'=\la,
\eeq
then the transformed metric remains a solution.
{Since we}
have the time rescaling freedom{,}
{we may} write the
non-relativistic solution {of the low-energy \Ho~gravity
theory (\ref{horava})} in the following form
\begin{equation}
  d\tilde{s}^2=-N^2  dt^2+g_{ij}\left(dx^i+ \gamma N^i
  dt\right)\left(dx^j+ \gamma N^j\right), \label{finalscale}
\end{equation}
where $\gamma = \pm \sqrt{\kappa\xi}$.

Note that different ADM representations of
the Kerr metric under
the
$\gamma$ scaling, {\it e.g.,} Boyer-Lindquist
{\it vs} Kerr-Schild, will, in general, result in
inequivalent spacetimes. This follows from
the fact that the
deformation acts {only} on the foliation rather than
{on} the {full} spacetime metric.

Indeed, the four-dimensional Ricci scalar satisfies the
Gauss--Codazzi relation
\begin{equation}
  R^{(4)}= R-(K^{2}-K_{ab}K^{ab})-2
  \nabla_{\nu}[n^{\sigma }\nabla_\sigma-n^{\nu}\nabla_\sigma n^{\sigma }].
  \label{4D_curvature}
\end{equation}
Here, the {$\gamma$} scaling does not act on
three-dimensional scalar $R$, whereas
the extrinsic curvature and the unit normal $n^{\mu }$ transform
nontrivially. Since the different representations of
the Kerr spacetime have
different embeddings ($g_{ij}, K_{ij}, n^{\mu}$) of the
spatial hypersurfaces, they generally yield different four-dimensional
curvature invariants after the transformations,
although these invariants are the same
before the transformations.

On the other hand, the ADM mass and angular momentum of the deformed
solutions will
scale identically across the different representations
provided they agree before the transformations. To determine this scaling {in
  the conserved charges, we consider the recently proposed
  covariant formalism discussed in Ref.
  \cite{Deve:2021}, in the absence of $a_i$ extensions,
  for simplicity \footnote{The result is
    the same as that obtained with the standard
    canonical approach, which is based on the
    boundary terms for the differentiable action
\cite{Deve:2021}.}}. By construction, the conserved charge
is obtained from the covariant conservation law together with Stokes'
theorem
\beq
Q(\xi)&=&\int d^{3}x \sqrt{g}\,n_{\mu}{J}^{\mu}\\
&={}& \int d^{3}x (\xi^{0}{\cal L}+\partial_{i} {\cal U}^{0 i}), \label{Q}
\eeq
where
\beq
{\cal L}&=&\frac{1}{\kappa}(K_{ij}K^{ij}-\lambda K^{2})+\xi R -2 \Lambda ,\\
{\cal U}^{0
j}&={}&\frac{2}{\kappa}\left(\xi^{0}N_{i}+\xi_{i}\right)\left(K^{ij}-\lambda
g^{ij}K\right).
\eeq
Decomposing this charge into temporal and spatial parts
\begin{equation}
  Q(\xi)=\gamma^2 Q_0(\xi^{0})+ \gamma Q_i(\xi^{i}),
  \label{charge}
\end{equation}
where $Q_0(\xi^{0})$ denotes the contribution associated with
the zeroth component of the Killing vector and $Q_i(\xi^{i})$ is the spatial
contribution. For the solutions we will consider{,} these will
correspond to ADM mass and angular momentum{,} respectively.
Thus, under the $\gamma$-scaling transformations, the
mass $M=\gamma^2 Q_0$ and angular momentum
$J=\gamma Q_i$
remain identical between the two representations if
their respective values are identical
before the transformations, even though $M$ and $J$
have different scaling weights.

\section{Examples}

In our previous work \cite{Deve:2024}, we
found the Kerr-like rotating black-hole solutions in Boyer-Lindquist coordinates
\cite{Kerr:1963}
{by considering the mass-dependent ansatz with three undetermined
functions, starting from}
the {massless} rotating solution {\cite{Gibb:1977,Park:2001,Gibb:2017}, which is
  an
  exact {\it vacuum} solution of
{\it full} \Ho~gravity \cite{Park:2023}}. In this section, we
consider two illustrative examples in different coordinate systems{,
  using the simple scaling transformations 
discussed in the previous section}.

The first example is an
asymptotically flat Doran-like rotating
solution
with unit lapse.
By simply scaling the shift function of
{
  the Doran metric in GR \cite{Dora:1999}, or even in
  the presence of the
  $a_i$ extensions, with $\gamma=\pm \sqrt{\kappa
  \xi},~\la=1$, and $\La=0$,
  we
  obtain the following
  solution in
  low-energy \Ho~gravity,
  {
    \begin{eqnarray}
      ds^2&=&-N^2 dt^2+ \rho^2
      d \theta
      ^{2}+(r^{2}+a^{2})\sin^{2}\theta~ d \phi ^{2} \no \\
      &&+\left[
        \frac{\rho }{\sqrt{r^{2}+a^{2}}}(dr+N^r dt)-\frac{a
        \sqrt{2mr} \sin^{2}\theta}{\rho}d \phi
      \right]^{2} \no \\
      &=&-\left(1- \gamma^2\,\frac{ 2mr}{\rho^2} \right)
      dt^2+\frac{\rho^2}{(r^2+a^2)} dr^2 +\rho^2 d \theta
      ^{2}+\frac{\Sigma^2 \sin^{2}\theta}{\rho^2} d \phi^2 \no \\
      &&-2a\sqrt{\frac{2mr}{r^2+a^2}}~\sin^{2}\theta~ dr~ d
      \phi+2N_r~ dt~ dr +2 N_\phi~ dt~ d\phi,
      \label{Doran-like}
    \end{eqnarray}
    where
    \begin{eqnarray}
      &&\rho^2= r^{2}+a^2 \cos^{2}\theta,~~\Sigma^2 =(r^2+a^2)
      \rho^2+ 2 m r a^2 \sin^2 \theta, \no \\
      &&N^2=1,~~ N^r=\frac{\gamma \,\sqrt{2mr (r^2+a^2)}}{\rho^2},\no \\
      &&N_r=\gamma\, \sqrt{\frac{2mr}{(r^2+a^2)}},~~
      N_{\phi}=-\frac{\gamma\, 2mr a \sin^2 \theta}{\rho^2},
      \label{Doran-convention}
    \end{eqnarray}
  }
  with $N^{\theta}=N^{\phi}=0$.
  The {four-}dimensional curvature scalar of this solution is
  \begin{equation}
    R^{(4)}
    =
    \frac{2 m ra^2(1-\gamma^2)
    \left(3\sin^2\theta-2\right)}
    {\rho^6}.
    \label{R_Doran-like}
  \end{equation}
  In the previous section, we have argued that the geometry of
  the transformed
  solution with the $\gamma$ scaling will be different from
  its seed metric. Here, we see this
  explicitly:
  even the asymptotically
  flat solution now acquire{s} a
  non-trivial curvature scalar{, which also exhibits
  the usual ring singularity at $\rho=0$, {\it i.e.}, $r=0,~\theta=\pi/2$}.


  The second solution we consider is the {asymptotically (Anti)de Sitter
    Kerr-Schild-like solution, obtained
    from the Kerr-(Anti)de Sitter metric in Kerr-Schild form as the
    seed solution with
    \cite{Gibb:2004},
    \begin{equation}
      ds^2= d\bar{s}^{2}+\frac{2mr}{\rho^2}(k_{\mu}dx^{\mu })^{2}+
      (1-\gamma^2) \frac{2mr \left(1-\frac{\Lambda }{3} r^{2} \right)
      \Delta^2_{\theta}}{\Xi \Upsilon^{2}} dt^2{,}
      \label{KS-like}
    \end{equation}
    with the {\it zero-mass} Kerr-(Anti)de Sitter
    background metric
    \begin{equation}
      d\bar{s}^2= -\frac{(1-\frac{\Lambda }{3} r^{2}) \Delta_\theta
      }{\Xi} d t^2+ \frac{\rho^{2} }{(1-\frac{\Lambda }{3}
      r^{2})(r^{2}+a^{2})} d r^{2}+\frac{\rho^{2}}{\Delta_\theta} d \theta ^{2}
      +\frac{(r^{2}+a^{2}) \sin^{2}\theta }{\Xi} d \phi ^{2}{,}
      \label{KS-like-zero}
    \end{equation}
    and one-form
    \begin{eqnarray}
      k_{\mu} dx^{\mu}=\gamma \frac{\Delta_{\theta}}{\Xi}dt
      +\frac{\rho^2}{(1-\frac{\Lambda }{3} r^{2})(r^{2}+a^{2})}dr
      -\frac{a \sin^{2}\theta }{\Xi} d \phi,
    \end{eqnarray}
    where
    \begin{eqnarray}
      \rho^2&=& r^{2}+a^{2}\cos ^{2}\theta ,~\Xi=1+\frac{\Lambda }{3} a^2,~
      \Delta_\theta = 1+\frac{\Lambda }{3} a^{2}\cos ^{2}\theta, \no \\
      \Upsilon^{2}&=&\rho^2 \Xi \left(1-\frac{\Lambda }{3}
      r^2 \right)+2mr~ \Delta_{\theta}.
    \end{eqnarray}
    Note that the \Ho~scaling $\ga$ breaks the nullness of $k_{\mu}$,
    $\bar{g}^{\mu \nu} k_{\mu}k_{\nu}=(1-\ga^2)
    \De_{\theta}/(1-\frac{\Lambda }{3} r^2 )\Xi,~{g}^{\mu \nu}
    k_{\mu}k_{\nu}=(1-\ga^2) \rho^2
    \De_{\theta}/\Upsilon^{2}$, unless $\ga^2=1$.
    On the other hand, when expressed in the ADM form, the metric
    (\ref{KS-like}) can also be written as
    \begin{eqnarray}
      ds^2&=&-N^2 dt^2+ \frac{\rho^2 \left[(r^2+a^2) (1-\frac{\Lambda
      }{3} r^{2})+2mr\right]}{(r^2+a^2)^2 (1-\frac{\Lambda }{3}
      r^{2})^2} (dr+N^r dt)^2 +\frac{\rho^2}{\Delta_{\theta}} d
      \theta ^{2} \no \\
      &&+\frac{\Sigma^2 \sin^{2}\theta}{\rho^2 \Xi^2} (d
      \phi+N^{\phi} dt)^2-\frac{4amr \sin^2 \theta}{\Xi ~(r^2+a^2)
      (1-\frac{\Lambda }{3} r^{2})} (dr+N^r dt)(d \phi+N^{\phi} dt),
      \label{KS-like_ADM}
    \end{eqnarray}
    where
    \beq
    &&{N^2=\f{\rho^2  (1-\frac{\Lambda }{3} r^{2})^2
      \De_{\theta}}{\Upsilon^{2}},~~N^r=\ga \f{2mr
    (1-\frac{\Lambda }{3} r^{2}) \De_{\theta}}{\Upsilon^{2}},}\no \\
    &&N^{\phi}=\f{-a}{r^2+a^2}N^r, ~~\Sigma^2 =(r^2+a^2) \rho^2 \Xi +
    2 m r a^2 \sin^2 \theta.
    \eeq
  }
  In order to compare the resulting geometry with the Doran{-like}
  representation of {the rotating black-hole metric
  (\ref{Doran-like}),}
  we set
  {$\Lambda =0$}, for which the seed geometry reduces to the
  Ricci-flat Kerr metric.
  The corresponding four-dimensional Ricci scalar {of
  the Kerr-Schild-like metric (\ref{KS-like}) or (\ref{KS-like_ADM})} is
  \begin{equation}
    R^{(4)}
    =\frac{2(\gamma^2-1)m^2\,
      \mathcal{P} 
    }
    {\rho^6\left(\rho^2+2mr\right)^3},
  \end{equation}
  where
  \begin{eqnarray}
    \mathcal{P}
    &=& \frac{1}{2}a^{8}\cos^{6}\theta
    \left(3\cos 2\theta+5\right)
    +a^{6}r\sin^{2}\theta\cos^{4}\theta
    \left(2m+3r\right)
    + a^{4}r^{3}\cos^{2} \theta \left[
      \sin^{2}\theta(15r-4m)  \right. \no \\
      &&\left.    +8m-12r
    \right] +a^{2}r^{4}
    \left[
      16m^{2}\cos^{2}\theta
      +r\left(
        \sin^{2}\theta(2m+9r)-8r
      \right)
    \right]
    -8mr^{7}.
  \end{eqnarray}
  This result, 
  which differs from (\ref{R_Doran-like}), is a direct manifestation
  of the effect of embedding
  dependence
  discussed above: different types of seed metrics
  will
  generally result in
  the different four-dimensional
  Ricci scalar $R^{(4)}$ and
  {$\gamma$-scaled} geometries {in \Ho~gravity}, even
  though they are the same in GR.

  The final task we undertake is to calculate
  the mass and angular momentum of these solutions.
  Using the conserved charge formalism {\cite{Deve:2021}} described
  above,
  we obtain
  \begin{equation}
    M=16 \pi \xi  m, \quad J= {\pm}16 \pi  a m \sqrt{\xi \kappa ^{-1}
    },
  \end{equation}
  which
  scale identically for the two
  metrics,
  because their $M$ and $J$ values are
  the same 
  in GR before transformations, in agreement
  with the general form, given in (\ref{charge}) with
  $\ka \equiv 16 \pi$.

  \section{New Singularities, Non-Killing Horizons, and Chronology Protection}

  The new rotating solutions
  found in the previous section have some peculiar properties
  compared to the previously obtained rotating black-hole solutions, as well as
  those of the Kerr solution in GR
  in Boyer-Lindquist coordinates.\\

  1. First,
  we note that, since the apparent gravitational symmetry of the
  \Ho~action (\ref{horava}) is the {\it foliation-preserving}
  diffeomorphism ({\DiffF}), the physical singularities are captured
  by {\DiffF}~-invariant curvatures, up to finite factors,
  \beq
  &&K\sim
  \f{\sqrt{\ka \xi m}}{\sqrt{
  r (r^2+a^2)} \rho^2},~K_{ij} K^{ij} \sim \f{
  \ka \xi m}{r (r^2+a^2) \rho^6},~
  R\sim\f{
  r m a^2}{{\rho^6}}, \no \\
  &&R_{ij} R^{ij} \sim \f{
  m^2 a^4}{(r^2+a^2) \rho^{12}},~
  K_{ij} R^{ij} \sim
  \f{ \sqrt{
    \ka \xi r (r^2+a^2)}~ m^{2} a^2
  }{\rho^{10}}{,}
  \label{3_curvature_Doran-like}
  \eeq
  for the Doran-like solution (\ref{Doran-like}), and
  \beq
  K\sim
  \f{\sqrt{\ka \xi m}}
  {\rho^3 {~\Upsilon^3}}, ~
  K_{ij} K^{ij} \sim \f{\ka \xi m^2}
  {\rho^6 ~\Upsilon^6},~
  R\sim\f{m^2}
  {\rho^6 ~\Upsilon^4}, ~R_{ij} R^{ij} \sim \f{m^2 }{\rho^{12} ~\Upsilon^8},~
  K_{ij} R^{ij}\sim
  \f{\sqrt{\ka \xi}~m^2}{\rho^{9} ~\Upsilon^7}{,}
  \label{3_curvature_KS-like}
  \eeq
  for the Kerr-Schild-like solution (\ref{KS-like}) or (\ref{KS-like_ADM}).
  Note that (\ref{3_curvature_Doran-like}) and
  (\ref{3_curvature_KS-like}) show the new
  singularities at
  $r=0$ (for arbitrary $\theta$) and
  $\Upsilon^2=\rho^2+2mr=0$, respectively, which include
  the usual {\it ring} singularity at $\rho^2=0$, {\it i.e.}, $r=0,
  \theta=\pi/2$. (Here, we consider only the asymptotically flat case
    with $\Lambda=0$ for simplicity, as in the previous section, since
  the main conclusions are unchanged even with $\Lambda \neq 0$.)

  In general, these
  new singularities in three-dimensional curvatures may not be
  present in the four-dimensional curvature invariants, since
  these
  can be canceled in four-dimensional quantities such as
  (\ref{4D_curvature}). Indeed,
  one can check that the four-dimensional curvature invariants in the
  Doran-like metric have only the usual ring singularity at $\rho^2=0$,
  \beq
  &&
  R^{(4)}\sim  \f{(\ka \xi-1)m a^2}{\rho^2},~R^{(4)}_{\mu \nu}R^{(4)
  \mu \nu}\sim \f{ (\ka \xi-1)^2m^2 a^4}{(r^2+a^2)\rho^{12}}, \no \\
  &&R^{(4)}_{\mu \nu \alpha \beta}R^{(4) \mu \nu \alpha \beta}\sim
  \f{(\ka \xi-1)m^2}{(r^2+a^2)\rho^{12}} + (\cdots) \f{m^2}{\rho^{12}}.
  \eeq
  On the other hand, in the Kerr-Schild-like metric, one can clearly
  see the sharp appearance of the new singularities for
  $\ka \xi \neq 1$, in contrast to the usual ring singularity of
  $\rho^2=0$ in the GR case $\ka \xi =1$,
  \beq
  &&R^{(4)}\sim  \f{(\ka \xi-1)m^2}{\rho^6 ~\Upsilon^6},~R^{(4)}_{\mu
  \nu}R^{(4) \mu \nu}\sim  \f{(\ka \xi-1)^2 m^4}{{\rho^{12}~
  \Upsilon^{12}}}, \no \\
  &&R^{(4)}_{\mu \nu \alpha \beta}R^{(4) \mu \nu \alpha \beta}\sim
  \f{(\ka \xi-1) m^2}{\rho^{12}~\Upsilon^{12}} + (\cdots) \f{m^2}{\rho^{12}}.
  \eeq
  Here, $(\cdots)$ are the same factors as in GR.

  Then, one
  find that
  the Doran-like metric (\ref{Doran-like}) and the
  Kerr-Schild-like metric (\ref{KS-like_ADM}) are
  regular everywhere,
  except at
  the {new} curvature singularities
  $r=0$ and $\Upsilon^2=0$,
  respectively
  \footnote{For the asymptotically de Sitter metric with $\Lambda>0$,
    the metric (\ref{KS-like_ADM})
    shows another singularity at $r=\sqrt{3/\Lambda}$ but it is a
    coordinate singularity
    located
  beyond the cosmological horizon.}. \\

  2. The event horizon is a {\it null} hypersurface, generated by
  null geodesics that are trapped within the surface. For stationary
  axisymmetric solutions with the Killing vectors $\partial_t$ and
  $\partial_{\phi}$ as in our examples, let us consider a
  hypersurface
  defined by $\zeta (r,\theta) \equiv r-r_{H} (\theta)=0$
  to which
  Killing vectors are tangent, assuming a unique
  horizon radius $r_H (\theta)$ for any angle $\theta$
  \cite{Thor:2006,Joha:2013}.

  The condition for
  a null hypersurface is given by
  \beq
  \left.\left( g^{rr}-2 g^{r \theta} r_H ' (\theta)+ g^{\theta
  \theta} {r_H '}^2 (\theta)\right)\right|_{r=r_H}=0{,}
  \label{null_hyper}
  \eeq
  from the {\it nullity} of the normal vector,
  $\hat{n}^{\mu}\hat{n}_{\mu}=0$ with $\hat{n}_{\mu}=\nabla_{\mu}
  \zeta=(0,1,r_H ',0)$.
  Evaluating (\ref{null_hyper}) for the Doran-like metric
  (\ref{Doran-like}) and the Kerr-Schild-like metric
  (\ref{KS-like_ADM}), we find ($g^{r \theta}=0$)
  \beq
  {r_{H ({\rm Doran})}'}^2 (\theta)&=&\left.\left(-\Delta (r)
  +\f{(\ka \xi-1) (r^2+a^2) 2m r}{\rho^2}\right)\right|_{r=r_H} ,
  \label{null_Doran}\\
  {r_{H ({\rm KS})}'}^2 (\theta)&=&\left.\left(-\Delta (r) +\f{(\ka
  \xi-1) 4 m^2 r^2}{\Upsilon^2}\right)\right|_{r=r_H} ,  \label{null_KS}
  \eeq
  evaluated at $r=r_{H} (\theta)$, where $\De (r)=r^2+a^2-2mr$.

  For the GR limit $\ka \xi=1$, the null hypersurface is given by
  $\De(r)=0, r_H ' (\theta)=0$ with the usual Kerr
  horizon $r_H \equiv r_{\pm}=m \pm \sqrt{m^2-a^2}$, which is the
  same as in the Boyer-Linquist coordinates. However,
  for $\ka \xi \neq 1$, as long as the right-hand sides of
  (\ref{null_Doran}) and (\ref{null_KS}) are positive, the radii of
  the null hypersurfaces are $\theta$-dependent and
  can be obtained perturbatively, in powers of $\ep\equiv \ka \xi-1$,
  as $r=r_{H} (\theta)=r_{\pm}+\ep h_{\pm}(\theta)+{\cal O} (\ep^2)$ with
  \begin{equation}
    h_\pm^{\mathrm{Doran}}(\theta)
    =
    \frac{(r_\pm^2+a^2)^2}
    {(r_\pm-r_\mp)
    \left(r_\pm^2+a^2\cos^2\theta\right)},
    \\ \quad
    h_\pm^{\mathrm{KS}}(\theta)
    =
    \frac{(r_\pm^2+a^2)^2}
    {(r_\pm-r_\mp)
    \left(2r_\pm^2+a^2(1+\cos^2\theta)\right)}.
  \end{equation}
  (See Fig.1 for the full numerical integrations of the non-linear
  equations (\ref{null_Doran}) and (\ref{null_KS}).)

  \begin{figure}
    \includegraphics[width=8cm,keepaspectratio]{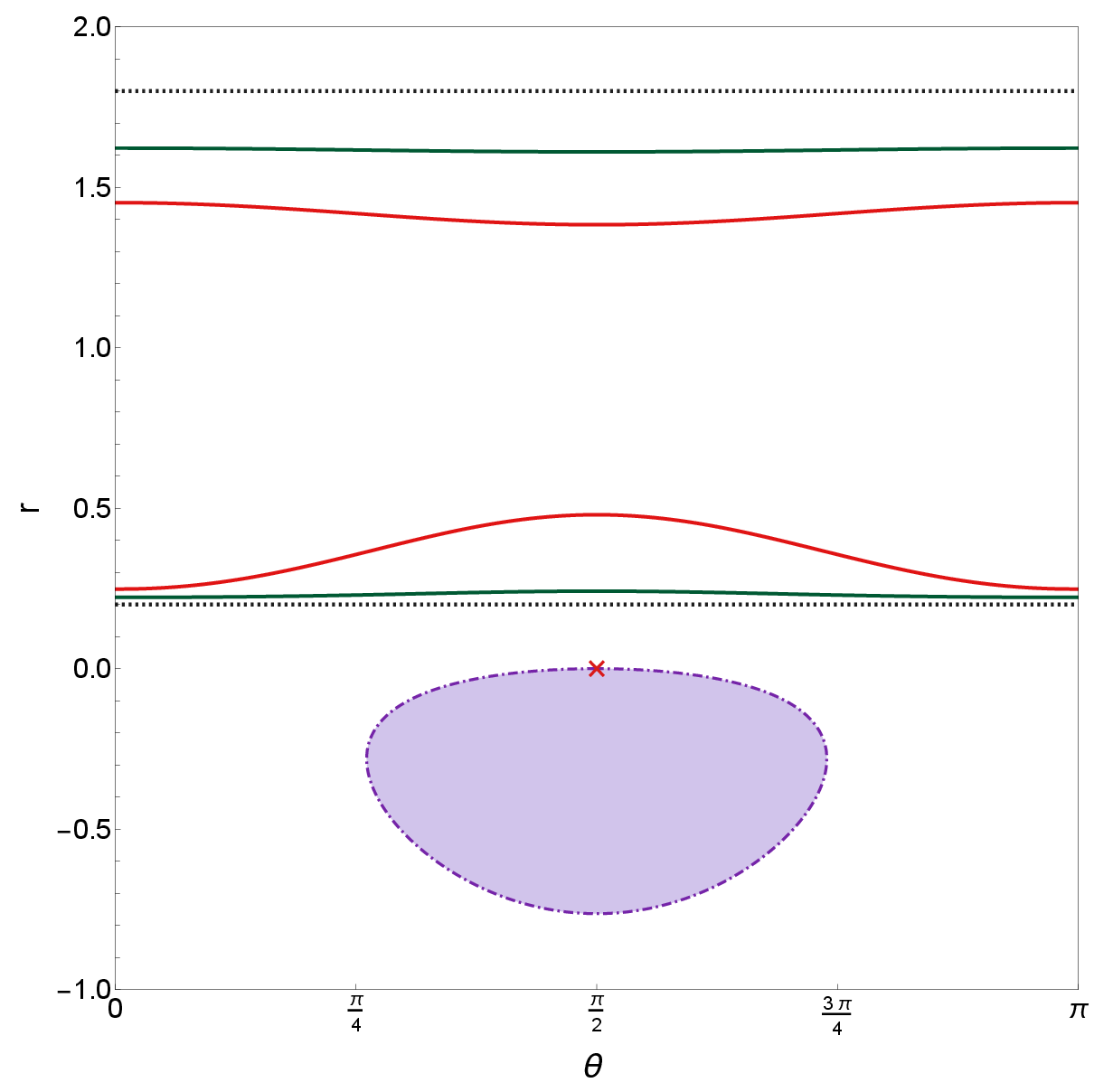}
    \caption{Numerical integrations of $r_H(\theta)$ for the null
      hypersurface equations (\ref{null_Doran}) and (\ref{null_KS})
      that smoothly converge
      to the Kerr horizons $r_{\pm}$ in the GR limit $\ka \xi=1$.
      The outermost (upper and lower) {\it constant} dotted-lines
      denote the Kerr
      horizons $r_+$ and $r_-$ in GR, respectively, and the
      $\theta$-dependent $r_H (\theta)$ for the Doran-like metric and
      the Kerr-Schild-like metric
      lie in their respective interior
      regions, {\it i.e.} $r_{+{\rm Doran}}<r_{+ {\rm
      KS}}<r_{+}$ and $r_{-}<r_{-{\rm
      KS}}<r_{-{\rm Doran}}$ for $m=1,a=0.6, \ka
      \xi=0.85$. The red
      cross denotes the usual ring singularity of $\rho^2=0$ located
      at $(r,\theta)=(0,\pi/2)$ and the closed purple region in the
      region $r<0$ denotes $g_{\phi \phi}<0$, determined by
      $\Si^2<0$, with its boundary at $g_{\phi \phi}=0$, {\it i.e.,} $\Si^2=0$.
    }
    \label{fig:CTC}
  \end{figure}

  An important property of the nonconstant-radius null
  hypersurfaces is that they are not Killing
  horizons, since the Killing vector ${\bf
  \xi}=\pa_t+\Om_{\phi} \pa_{\phi}$ with a constant $\Om_{\phi}$
  cannot be null
  but spacelike $\xi \cdot
  \xi|_{r=r_H}=g_{\phi \phi} (\Om|_{\phi}+g_{t \phi}/g_{\phi
  \phi})^2+{r_H'}^2/g_{\phi \phi}>0$ at the null
  hypersurfaces $r=r_H(\theta)$ for $g_{\phi \phi}>0$.
  Black holes with non-Killing horizons have been found in various
  contexts \cite{Figu:2012,Fisc:2012, Joha:2013, Anso:2020,BenA:2020,
  Adam:2021}. In our solutions, non-Killing horizons are
  essential consequences
  of
  low-energy Lorentz violation for
  $\ka \xi \neq 1$ in the {{\it horizon-penetrating}}
  Doran-like and Kerr-Schild-like coordinates, {
  contrary} to the Killing horizons in the previously {obtained}
  rotating black holes in the Boyer-Linquist
  coordinates, which have the coordinate
  singularities at the horizons $r_{\pm}$.\\

  3. In the Kerr metric,
  or {in its alternative
  forms} in the Doran
  \cite{Dora:1999} {and}
  the Kerr-Schild
  \cite{Kerr:1965, Gibb:2004} {coordinates}, there exists
  a {causality-violating region} ${\cal V}=\{ g_{\phi \phi}<0\}$,
  located where
  $\Sigma^2=(r^2+a^2) \rho^2+2ma^2 {\rm sin}^2\theta<0$ for $mr<0$
  and contains closed-timelike curves (CTCs) with a timelike vector
  $\pa_{\phi}$ \cite{Cart:1968}. The
  causality-violating region ${\cal V}$ in the
  Doran-like metric (\ref{Doran-like}) and the Kerr-Schild-like
  metric (\ref{KS-like_ADM}) are identical
  as in the Doran or Kerr-Schild metric because $g_{\phi
  \phi}$ has
  the same form in both metrics. However, the
  causality-violating region is inaccessible
  because of
  the new curvature singularities at $r=0$ or $\Upsilon^2=0$ for the
  Doran-like metric or the Kerr-Schild-like metric, respectively,
  which prevent
  one from reaching
  or coming out of it (Fig. 2). Hence, the region ${\cal
  D}_+=\{r>0\}$ for the Doran-like metric and the region ${\cal
  M}_+=\{\Upsilon^2>0\}$ for the Kerr-Schild-like metric are
  {\it chronological.}
  Thus, we find a {\it chronology
  protection} \cite{Hawk:1991} in our Doran-like
  and Kerr-Schild-like rotating solutions due to new
  curvature singularities outside the
  causality-violating region when there is the Lorentz
  violation for
  $\ka \xi \neq  1$, as in the previously found Kerr-like metric \cite{Park:2025}.

  Moreover, the regular domains ${\cal D}_+$ and ${\cal M}_+$ are
  {\it stably causal} \cite{Hawk:1973} because $T=t$ is a time
  function, with
  an increasing $T$ defining
  the future direction, shown by $\nabla_{\mu}T \nabla^{\mu}T=-1$ for
  the Doran-like metric and $\nabla_{\mu}T
  \nabla^{\mu}T=-\Upsilon^2/\rho^2<0$ for the Kerr-Schild-like
  metric. Note that the Doran metric has a
  similar casualty property
  in
  its defining region $r > 0$, but there is no
  obstruction to the Kerr extension in the region $r \leq 0$,
  contrary to our Doran-like metric. The Kerr-Schild metric also has
  a
  similar causality property in
  the region $\Upsilon^2>0$, but there is no physical
  obstruction to reaching $\Upsilon^2 \leq 0$,
  containing the causality-violating region $\Si^2<0$ for $mr<0$.

  \begin{figure}
    \includegraphics[width=5.05cm,keepaspectratio]{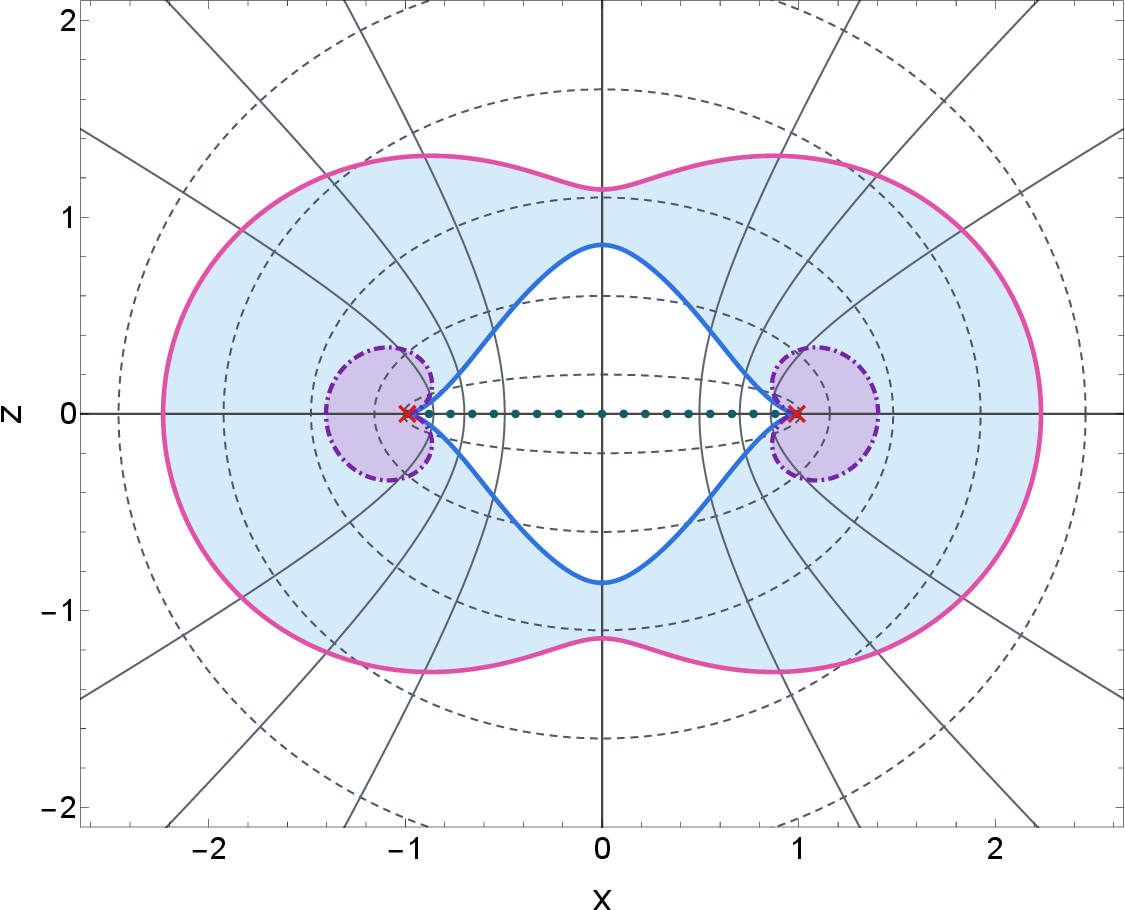}
    \quad
    \includegraphics[width=5.05cm,keepaspectratio]{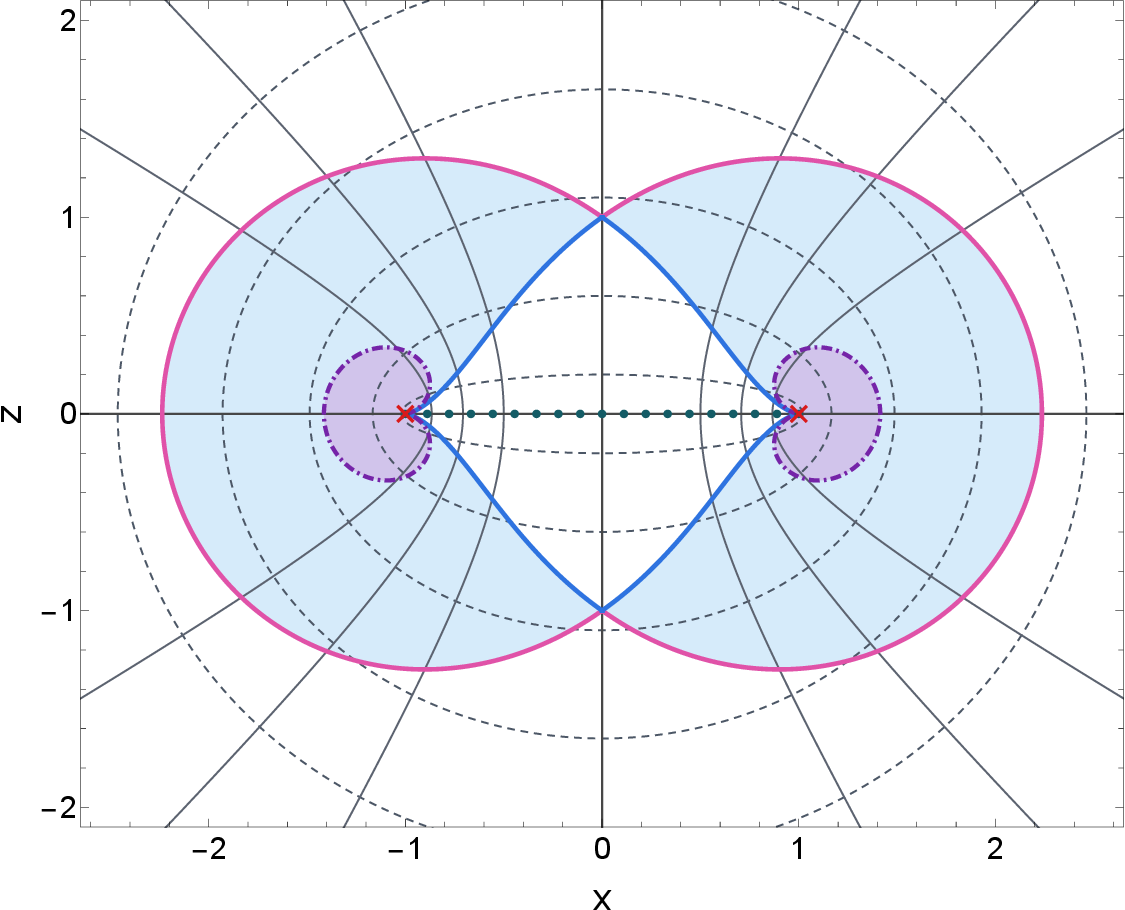}
    \quad
    \includegraphics[width=5.3cm,keepaspectratio]{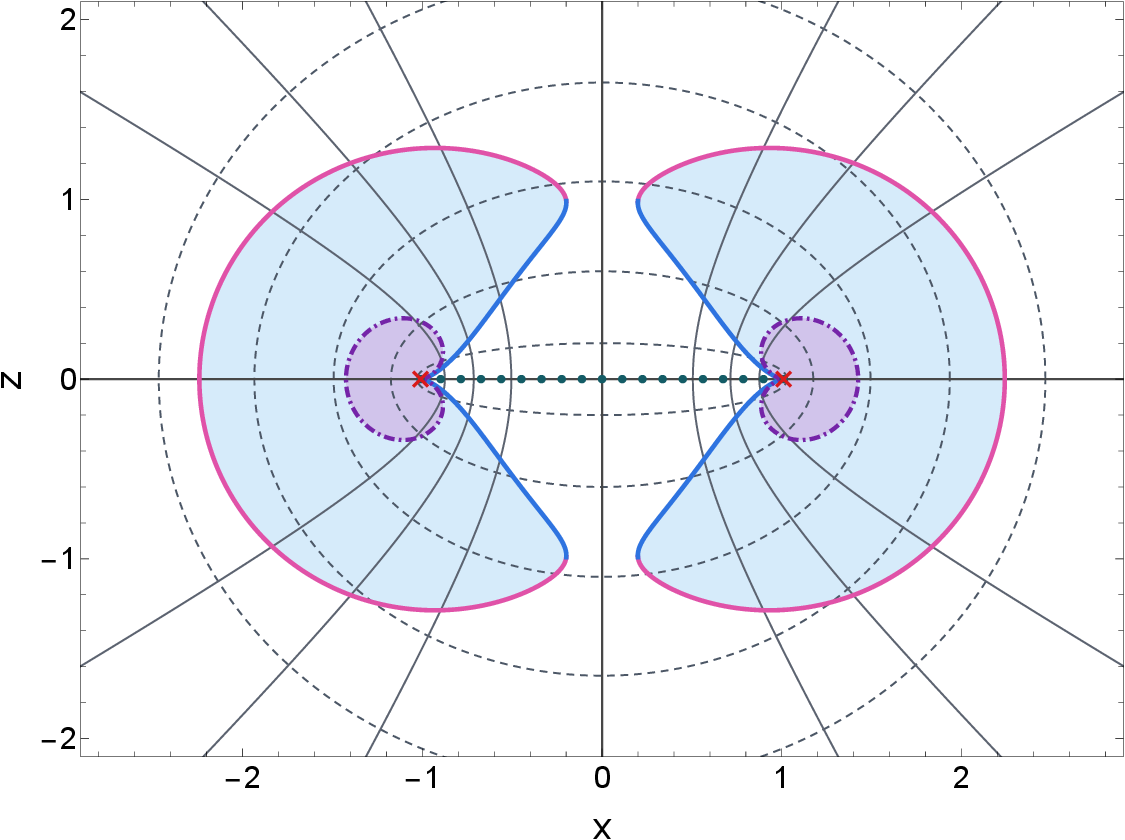}
    \quad
    \caption{Plot of singularities of $r=0$ (green dots) and
      $\Upsilon^2=0$ (pink and blue solid lines) for the Doran-like
      metric and the Kerr-Schild-like metric, respectively,
      together with
      the causality-violating region ${\cal
      V}=\{g_{\phi \phi} <0\}$ (purple region), determined by
      $\Sigma^2=(r^2+a^2) \rho^2+2m r a^2 {\rm sin}^2\theta<0$ in the
      region $r<0$ for $m>0$. The figure shows meridional sections in the $y=0$ plane of the Cartesian coordinate system $(x,y,z)$, with $x=\pm \sqrt{r^2+a^2} \rm{sin} \theta$ and $z=r \rm{cos}\theta.$
    }
    \label{fig:Singular}
  \end{figure}

  \section{Concluding Remarks}

  We have presented scaling transformations that generate new
  rotating black-hole solutions
  in the low-energy \Ho~gravity
  from relativistic solutions. We
  first obtained Doran-like and Kerr-Schild-like rotating solutions
  in \Ho~gravity
  as explicit examples and have shown
  they are
  horizon-penetrating  and have
  non-Killing horizons, new singularities and
  {
  chronology-protecting} properties.

  We remark that the Doran-like solution is the first
  {``projectable"}
  rotating black-hole
  solution and the Kerr-Schild-like solution is the first
  {``non-projectable"} rotating
  black-hole solution that is horizon-penetrating, in contrast to
  the previously obtained
  Kerr-like solutions in Boyer-Lindquist coordinates which
  have coordinate
  singularities at the Killing horizons \cite{Deve:2024}.
  It
  would be interesting to see whether the new rotating solutions
  could approximately admit the usual black hole
  thermodynamics or mimic Hawking's area theorem
  in the presence of
  the tiny but non-vanishing Lorentz violations and thus
  they could be consistent
  with
  the recent observational data \cite{LIGO:2025, Vang:2026}.
  It may be surprising that the chronology is protected in a
  natural
  way in
  non-relativistic gravity, rather than relativistic
  theories, but the stable casuality with the preferred
  time coordinate might be the core reason for this
  phenomena \cite{Isma:2026}.

  \section*{Acknowledgments}

  This work was supported by Basic Science Research Program through the National
  Research Foundation of Korea (NRF) funded by the Ministry of Education,
  Science and Technology (RS-2020-NR049598 (MIP)).

  \appendix
  \section{Field Equations}
  \beq
  {\cal H}&\equiv&\f{1}{\kappa}\left(K_{ij}K^{ij} -\lambda
  K^2\right)-\xi R +2 \La {-\si
    \left(\frac{1}{2}\frac{\nabla_i N \nabla^i N}{N^2}-\frac{\nabla_k
  \nabla^k N}{N} \right)=0}, \\
  {\cal H}^i&\equiv&{-}\f{2}{\ka}\nabla_j
  \left(K^{ji}-\lambda\,Kg^{ji}\right)=0\, ,\\
  E_{ij}&\equiv&\frac{1}{\kappa}\left( E_{ij}^{(1)}-\lambda E_{ij}^{(2)} \right)
  +\xi E_{ij}^{(3)}+\La N g_{ij}+\frac{\si}{2} E_{ij}^{(4)}=0 .
  \eeq
  where
  \bea
  E_{ij}^{(1)}&=& N_i \nabla_k K^k{}_j + N_j\nabla_k K^k{}_i -K^k{}_i
  \nabla_j N_k-
  K^k{}_j\nabla_i N_k - N^k\nabla_k K_{ij}\no\\
  && - 2N K_{ik} K_j{}^k
  -\frac{1}{2} N K^{k\ell} K_{k\ell}\, g_{ij} + N K K_{ij} + \dot K_{ij}
  \,,\no \\
  E_{ij}^{(2)}&=& \frac{1}{2} NK^2 g_{ij}+ N_i \pa_j K+
  N_j \pa_i K- N^k (\pa_k K)g_{ij}+  \dot K\, g_{ij}\,,\no\\
  E_{ij}^{(3)}&=&N\Big(R_{ij}- \frac{1}{2}R g_{ij}
  {\Big)}-(
  \nabla_i\nabla_j-g_{ij}\nabla_k\nabla^k)N\,,\no\\
  { E_{ij}^{(4)}} &=& {\frac{1}{N}
  \Big(-\frac{1}{2}g_{ij} \nabla_k N \nabla^k N +\nabla_i N \nabla_j N \Big). }
  \eea
  \newcommand{\J}[4]{#1 {\bf #2} #3 (#4)}
  \newcommand{\andJ}[3]{{\bf #1} (#2) #3}
  \newcommand{\AP}{Ann. Phys. (N.Y.)}
  \newcommand{\MPL}{Mod. Phys. Lett.}
  \newcommand{\NP}{Nucl. Phys.}
  \newcommand{\PL}{Phys. Lett.}
  \newcommand{\PR}{Phys. Rev. D}
  \newcommand{\PRL}{Phys. Rev. Lett.}
  \newcommand{\PTP}{Prog. Theor. Phys.}
  \newcommand{\hep}[1]{ hep-th/{#1}}
  \newcommand{\hepp}[1]{ hep-ph/{#1}}
  \newcommand{\hepg}[1]{ gr-qc/{#1}}
  \newcommand{\bi}{ \bibitem}


\begin{thebibliography}{999}
    \bibitem{LIGO:2016}
    B.~P.~Abbott \textit{et al.} [LIGO Scientific and Virgo],
    Phys. Rev. Lett. \textbf{116},
    061102 (2016).

    \bibitem{EHT:2019}
    K.~Akiyama \textit{et al.} [Event Horizon Telescope],
    Astrophys. J. Lett. \textbf{875}, L1 (2019).

    \bibitem{DeWi:1967}
    B.~S.~DeWitt,
    Phys.\ Rev.\  {\bf 160}, 1113 (1967).

    \bibitem{Hora:2009} P.~Ho\v{r}ava,
    Phys.\ Rev.\  D {\bf 79}, 084008 (2009).

    \bibitem{Blas:2009}
    D.~Blas, O.~Pujolas and S.~Sibiryakov,
    Phys. Rev. Lett. \textbf{104}, 181302 (2010).

    \bibitem{Park:2009}
    M.~I.~Park,
    Class. Quant. Grav. \textbf{28}, 015004 (2011).

    \bibitem{Deve:2021}
    D.~O.~Devecioglu and M.~I.~Park,
    Phys. Rev. D \textbf{108}
    064030 (2023).

    \bibitem{Deve:2024}
    D.~O.~Devecioglu and M.~I.~Park,
    Eur. Phys. J. C \textbf{84},
    852 (2024).

    \bibitem{Kerr:1963}
    R.~P.~Kerr,
    Phys.\ Rev.\ Lett.\  {\bf 11}, 237 (1963).

    \bibitem{Gibb:1977}
    G.~W.~Gibbons and S.~W.~Hawking,
    Phys. Rev. D \textbf{15}, 2738
    (1977).

    \bibitem{Park:2001}
    M.~I.~Park,
    Nucl. Phys. B \textbf{634}, 339
    (2002).

    \bibitem{Gibb:2017}
    G.~W.~Gibbons and M.~S.~Volkov,
    Phys. Rev. D \textbf{96},
    024053 (2017).

    \bibitem{Park:2023}
    M.~I.~Park and H.~W.~Lee,
    ``Zero-Mass Rotating Spacetimes in Four-Dimensional Horava Gravity,''
    [arXiv:2309.13859 [hep-th]].

    \bibitem{Dora:1999}
    C.~Doran,
    Phys. Rev. D \textbf{61}, 067503 (2000).

    \bibitem{Gibb:2004}
    G.~W.~Gibbons, 
    H.~Lu, D.~N.~Page and C.~N.~Pope,
    J. Geom. Phys. \textbf{53}, 49
    (2005), Appendix {\bf E}.

    {\bibitem{Thor:2006}
      J.~Thornburg,
      Living Rev. Rel. \textbf{10}, 3 (2007).

      \bibitem{Joha:2013}
      T.~Johannsen,
      Phys. Rev. D \textbf{87},
      124017 (2013).

      \bibitem{Figu:2012}
      P.~Figueras and T.~Wiseman,
      Phys. Rev. Lett. \textbf{110}, 171602 (2013).

      \bibitem{Fisc:2012}
      S.~Fischetti, D.~Marolf and J.~E.~Santos,
      Class. Quant. Grav. \textbf{30}, 075001 (2013).

      \bibitem{Anso:2020}
      T.~Anson, E.~Babichev, C.~Charmousis and M.~Hassaine,
      JHEP \textbf{01}, 018 (2021).

      \bibitem{BenA:2020}
      J.~Ben Achour, H.~Liu, H.~Motohashi, S.~Mukohyama and K.~Noui,
      JCAP \textbf{11}, 001 (2020).

      \bibitem{Adam:2021}
      A.~Adam, P.~Figueras, T.~Jacobson and T.~Wiseman,
      Class. Quant. Grav. \textbf{39},
      125001 (2022).

      \bibitem{Kerr:1965}
      R.~P.~Kerr and A.~Schild,
      Gen. Rel. Grav. \textbf{41},
      2485 
      (2009).

      \bibitem{Cart:1968}
      B.~Carter,
      Phys. Rev. \textbf{174}, 1559 
      (1968).

      \bibitem{Hawk:1991}
      S.~W.~Hawking,
      Phys. Rev. D \textbf{46}, 603 
    (1992).}

\bibitem{Park:2025}
M.~I.~Park and H.~Maeda,
Phys. Lett. B \textbf{872}, 140063 (2026).

    \bibitem{Hawk:1973}
    S.~W.~Hawking and G.~F.~R.~Ellis,
    ``The Large Scale Structure of Space-Time,''
    Cambridge University Press, 2023.

    \bibitem{LIGO:2025}
    A.~G.~Abac \textit{et al.} [LIGO Scientific, Virgo and KAGRA],
    Phys. Rev. Lett. \textbf{135},
    111403 (2025).

    \bibitem{Vang:2026}
    S.~Vangstein and C.~J.~Moore, ``On the Interpretation of Tests of
    the Horizon Area Law,''
    [arXiv:2610.36162 [gr-qc]].

    \bibitem{Isma:2026}
    M.~Ismail and D.~Mattingly, ``Metric--affine gravity with
    dynamical chronology protection,''
    [arXiv:2604.08511 [gr-qc]].

  \end{thebibliography}
  \end{document}